\documentclass[doublespacing]{elsart}
\usepackage{amssymb}

\usepackage{amsmath}

\journal{Journal of Theoretical Biology}
\input{tcilatex}
\begin{document}

\begin{frontmatter}
\title{The effects of distributed life cycles on the dynamics of viral infections}
\author[label1]{Daniel Campos\corauthref{cor}},
\corauth[cor]{Corresponding author}
\ead{Daniel.Campos@uab.es}
\author[label2]{Vicen\c{c} M\'{e}ndez} and 
\author[label1]{Sergei Fedotov}

\address[label1]{School of Mathematics, Department of Applied Mathematics. The
University of Manchester, Manchester M60 1QD, UK.}

\address[label2]{Grup de F\'{\i}sica Estad\'{\i}stica. Departament de F\'{\i}sica.
Universitat Aut\`{o}noma de Barcelona, 08193 Bellaterra (Barcelona) Spain}

\newpage

\begin{abstract}
We explore the role of cellular life cycles for viruses and host cells in an infection process. For this purpose, we derive a generalized version of the basic model of virus dynamics (Nowak, M.A., Bangham, C.R.M., 1996. Population dynamics of immune responses to persistent viruses. Science 272, 74-79) from a mesoscopic description. In its final form the model can be written as a set of Volterra integrodifferential equations. We consider the role of age-distributed delays for death times and the intracellular (eclipse) phase. These processes are implemented by means of probability distribution functions. The basic reproductive ratio $R_0$ of the infection is properly defined in terms of such distributions by using an analysis of the equilibrium states and their stability. It is concluded that the introduction of distributed delays can strongly modify both the value of $R_0$ and the predictions for the virus loads, so the effects on the infection dynamics are of major importance. We also show how the model presented here can be applied to some simple situations where direct comparison with experiments is possible. Specifically, phage-bacteria interactions are analysed. The dynamics of the eclipse phase for phages is characterized analytically, which allows us to compare the performance of three different fittings proposed before for the one-step growth curve.
\end{abstract}


\newpage

\end{frontmatter}

\section{Introduction}

The interactions between viruses and cells in an infection process can be
seen as an ecological system within the infected host. The mathematical
description of these systems has attracted increasing interest in the last
years (Wodarz, 2006), especially concerning the characteristics of the
immune response to a viral attack. A decade ago, Nowak and Bangham (1996)
presented what has been called thereafter the Basic Model of Virus Dynamics
(BMVD). This model has become quite popular among theorists and
experimentalists (see Nowak and May (2000) and Perelson (2002) for some
understanding reviews). The interplay between the BMVD and the effect of an
immune response has proved useful to describe the dynamics of chronic HIV
infections (Perelson, 2002). Furthermore, it has provided interesting
results regarding topics as the performance of drug therapies (Bonhoeffer et
al., 1997; Wodarz and Nowak, 1999), lymphocyte exhaustion (Wodarz et al.,
1998), etc.

The BMVD describes the time evolution of non-infected cells ($X$), infected
cells ($Y$) and viruses ($V$) by the system of equations 
\begin{eqnarray}
\frac{\text{d}X}{\text{d}t} &=&\lambda -\delta X-\beta XV  \nonumber \\
\frac{\text{d}Y}{\text{d}t} &=&\beta XV-aY  \nonumber \\
\frac{\text{d}V}{\text{d}t} &=&kY-\beta XV-uV.  \label{1}
\end{eqnarray}
The infection process is governed by the parameter $\beta $, which
determines the rate of successful contacts between the target cells and the
viruses. Mortality terms for the three species are considered with constant
death rates $\delta $, $a$ and $u$, respectively. The parameter $k$ measures
the rate at which virions are released from a single infected cell. Finally,
new target cells are produced by the host at a constant rate $\lambda $.

Despite the success achieved by the BMVD, it is clear that the model
described in (\ref{1}) is just a first approximation to the real underlying
process. Probably the strongest simplification made in the model is that it
assumes that the death rates are exponentially distributed (i.e.,
mortalities are considered as Markovian random processes) and therefore do
not take into account accurately the details of the cellular life cycles.
However, delays and structured life cycles are expected to play a very
significant role in the dynamics of viral infections. For example, the
infection process involves an intracellular phase of the virus, also known
as the eclipse phase, which is not explicitly considered in (\ref{1}). For
this reason, in the recent years some works have explored the effects of
constant and distributed delays in the BMVD, also in the case where an
immune response is considered. Herz et al., (1996) showed for the first time
the importance of delays in order to explain the virus loads observed in HIV
patients under drug treatment. This delayed model was later explored from a
more formal point of view by Tam (1999). Similar ideas, with different
expressions for the infection term, were considered by Culshaw and Ruan
(2000), Fort and M\'{e}ndez (2002) and Li and Wanbiao (1999). The effect of
distributed delays was explored for different models of virus dynamics by
Banks et al., (2003), Mittler et al., (1998) and Lloyd (2001). Finally, the
role of a delayed immune response has been the subject of extensive
research. Some examples are Buric et al., (2001), Canabarro et al., (2004),
Wang et al., (2007) and the references there in, which focused on the
chaotic patterns which can appear in these systems.

The papers mentioned above have helped us to understand how delays can
modify the cell-virus and virus-immune system dynamics. However, most of
those works focused on the case where only one of the processes (usually the
intracellular phase) is delayed. So, they do not considered the possibility
of different delays for each process, whose combined contributions could
modify the dynamical behavior of the system.

On the other hand, the introduction of delays in the virus dynamics has been
usually based on phenomenological (not always rigorous) arguments, without
providing a justification of the delayed equations proposed. Only in Banks
et al., (2003), Fort and M\'{e}ndez (2002) and Wearing et al., (2005) a more
formal discussion was provided. We stress that the implementation of delays
into dynamical models is sometimes tricky, as memory effects can lead to the
breakdown of hypothesis that are well established for Markovian processes.
In fact, there is currently a very active research on this subject from the
point of view of statistical mechanics (see, for example, Allegrini et al.,
2003, Allegrini et al., 2007, Rebenshtok and Barkai, 2007 and the references
therein). According to these ideas, a rigorous mathematical approach is
necessary to reach an accurate physical description of virus dynamics with
delays. Here, we propose a system of Volterra integrodifferential equations
which is a generalization of the BMVD. This system of equations is derived
from a mesoscopic approach where balance equations for each species ($X$, $Y$
and $V$) are considered explicitly. Mesoscopic descriptions as that
considered here (based on Continuous-Time Random Walk processes) have become
quite usual tools for the description of physical and biological processes.
At this stage, they have proved useful for the study of heat transport
(Emmanuel and Berkowitz, 2007), biological invasions (M\'{e}ndez et al.,
unpublished), tumor cell growth (Fedotov and Iomin, 2007), solute transport
in porous media (Berkowitz et al., 2000), earthquakes dynamics (Helmstetter
and Sornette, 2002), financial markets (Masoliver et al., 2006) and many
other. Here we will explore for the first time their application to the
field of virus dynamics.

Then, the aim of this paper is to use an integrodifferential approach to
show how distributed delays can strongly influence the predictions from the
BMVD. We find that the value of the basic reproductive ratio $R_{0}$ and the
values of the virus load can drastically change, in accordance with similar
conclusions found in Lloyd (2001) from the analysis of the intracellular
phase. Furthermore, the advantage of using such a general formalism as the
one proposed here is that different situations of interest can be analyzed
as particular cases of the model. According to this, we show how our model
can be used to fit and characterize the one-step growth (osg) curve observed
in phage-bacteria interactions. Three fittings proposed before by different
authors are compared. We find that, albeit the three approaches fit
reasonably well the osg curve, their predictions concerning the dynamics of
the eclipse phase are slightly different.

In the following, we show how a generalized version of the BMVD can be
obtained using a mesoscopic description. In Section 2 we present our model,
whose formal derivation is given in the Appendix for the sake of clarity. In
Section 3 we explore the equilibrium states and their stability, which let
us define the basic reproductive ratio $R_{0}$. After that, we consider
specific situations of special interest in virus dynamics. We consider the
effects of distributed delays in the phase eclipse (Section 4) and in the
mortalities for cells and viruses (Section 5). We also show how the model
derived in Section 2 works in the case of phages-bacteria dynamics (Section
6), and we provide some examples using experimental data extracted from the
literature. Finally, the main conclusions obtained from our study are
summarized in Section 7.

\section{The BMVD with distributed delays}

The model we consider here is depicted in Figure 1. It follows the same
scheme as the BMVD but some of the random processes (those indicated by the
dotted lines) are governed by their corresponding probability distribution
functions (PDF). So that, $\varphi _{X}(t)$ represents the probability that
a target cell $X$ dies at age $t$, with equivalent definitions for $\varphi
_{Y}(t)$ and $\varphi _{V}(t)$ for infected cells and viruses. Similarly,
the function $\phi (t)$ determines the dynamics of the eclipse phase: a cell
that becomes infected at time $t_{0}$ can release $\phi (t)$ viruses at time 
$t_{0}+t$.

The Volterra integrodifferential equations corresponding to the scheme in
Figure 1 read 
\begin{eqnarray}
\frac{\text{d}X(t)}{\text{d}t} &=&\lambda -\beta
X(t)V(t)-\int_{0}^{t}X(t-t^{\prime })\Psi _{X}(t^{\prime })\Omega
_{X}(t-t^{\prime },t)\text{d}t^{\prime }  \nonumber \\
\frac{\text{d}Y(t)}{dt} &=&\beta X(t)V(t)-\int_{0}^{t}Y(t-t^{\prime })\Psi
_{Y}(t^{\prime })\text{d}t^{\prime }  \nonumber \\
\frac{\text{d}V(t)}{\text{d}t} &=&-\beta X(t)V(t)+\int_{0}^{t}\beta
X(t-t^{\prime })V(t-t^{\prime })\phi (t^{\prime })\Phi _{Y}(t^{\prime })%
\text{d}t^{\prime }  \nonumber \\
&&-\int_{0}^{t}V(t-t^{\prime })\Psi _{V}(t^{\prime })\Omega _{V}(t-t^{\prime
},t)\text{d}t^{\prime }.  \label{11}
\end{eqnarray}
The formal derivation of this model in terms of a mesoscopic description is
provided in the Appendix. The functions $\Psi _{X}$, $\Psi _{Y}$, $\Psi _{V}$
are defined by their Laplace transforms (we denote the Laplace transform of
a function by the brackets $[$\textperiodcentered $]_{s}$ with the conjugate
variable $s$) 
\begin{equation}
\left[ \Psi _{X}\right] _{s}\equiv \frac{\left[ \varphi _{X}\right] _{s}}{%
\left[ \Phi _{X}\right] _{s}}\qquad \left[ \Psi _{Y}\right] _{s}\equiv \frac{%
\left[ \varphi _{Y}\right] _{s}}{\left[ \Phi _{Y}\right] _{s}}\qquad \left[
\Psi _{V}\right] _{s}\equiv \frac{\left[ \varphi _{V}\right] _{s}}{\left[
\Phi _{V}\right] _{s}},  \label{12}
\end{equation}
where $\Phi _{X}(t)\equiv \int_{t}^{\infty }\varphi _{X}(t^{\prime
})dt^{\prime }$ is the survival probability for the cells of age $t$.
Analogous definitions hold for $\Phi _{Y}$ and $\Phi _{V}$. According to (%
\ref{12}), the function $\Psi _{X}(t)$ can be interpreted as the
instantaneous death rate for a cell $X$ of age $t$. Then, the term $%
\int_{0}^{t}X(t-t^{\prime })\Psi _{X}(t^{\prime })\Omega _{X}(t-t^{\prime
},t)$d$t^{\prime }$ represents a generalized death term in which
age-distributed death rates are considered, and where $\Omega
_{X}(t-t^{\prime },t^{\prime })$ is the probability that a particle $X$ does
not become infected during the time interval $(t-t^{\prime },t)$. Similarly,
the term $\int_{0}^{t}\beta X(t-t^{\prime })V(t-t^{\prime })\phi (t^{\prime
})\Phi _{Y}(t^{\prime })$d$t^{\prime }$ represents the release of new
virions from those cells that became infected at time $t-t^{\prime }$,
provided that these cells have survived up to time $t$.

The system of equations (\ref{11}-\ref{12}) represents our generalization of
the BMVD to the case with distributed delays. An important conclusion from (%
\ref{11}) is that the density of infected cells $Y$ does not appear in the
equations for $X(t)$ and $V(t)$. It means that the formalism introduced here
allows us to reduce the BMVD to a 2-species model. We do not need to
consider explicitly the density $Y(t)$; the existence of the infected cells
is implicitly considered by means of the function $\Phi _{Y}$ appearing in
the equation for $V(t)$.

\section{Equilibrium states and their stability}

The equilibrium states of the model (\ref{11}) come from the analysis of the
fixed points of the system at $t\rightarrow \infty $. There are two possible
equilibrium states: the first one is the trivial, infection-free state,
given by 
\begin{equation}
(X_{eq},Y_{eq},V_{eq})=(\lambda \tau _{X},0,0).  \label{17}
\end{equation}
where we use $\tau _{i}=\int_{0}^{\infty }\Phi _{i}(t)$d$t$ to denote the
average lifetime of species $i$, with $i=X,Y,V$. The second state
corresponds to the case of a successful infection defined by 
\begin{eqnarray}
X_{eq}\int_{0}^{\infty }e^{-\beta X_{eq}t}\Phi _{V}(t)\text{d}t &=&\frac{%
\lambda \tau _{X}\int_{0}^{\infty }e^{-\beta \lambda \tau _{_{X}}t}\Phi
_{V}(t)\text{d}t}{R_{0}}  \nonumber \\
Y_{eq} &=&\lambda \tau _{Y}\beta V_{eq}\int_{0}^{\infty }e^{-\beta
V_{eq}t}\Phi _{X}(t)\text{d}t  \nonumber \\
\int_{0}^{\infty }e^{-\beta V_{eq}t}\Phi _{X}(t)\text{d}t &=&\frac{X_{eq}}{%
\lambda }  \label{18}
\end{eqnarray}
where the equations (\ref{5b},\ref{5c}) have been used, and we have defined 
\begin{equation}
R_{0}\equiv \beta \lambda \tau _{X}\left[ \int_{0}^{\infty }e^{-\beta
\lambda \tau _{_{X}}t}\Phi _{V}(t)\text{d}t\right] \left[ \int_{0}^{\infty
}\phi (t)\Phi _{Y}(t)\text{d}t\right] .  \label{19}
\end{equation}
As can be seen from (\ref{18}), it is not possible to give explicit
expressions for the equilibrium densities. However, it can be proved that
the infected state only has biological meaning ($Y_{eq}>0$ and $V_{eq}>0$)
if $R_{0}>1$. To see this, note that the condition $R_{0}>1$ applied to the
first equation of (\ref{18}) implies $X_{eq}<\lambda \tau _{X}$, which means
that the equilibrium density in the infected state is lower than in the
trivial state. Using that condition, it follows that the third equation in (%
\ref{18}) has necessarily a positive solution for $V_{eq}$. Hence, $R_{0}$
can be properly defined as the basic reproductive ratio, which is a key
parameter in epidemiology and virus dynamics in order to predict the
emergence of an infection (Anderson and May, 1991; Nowak and May, 2000). For 
$R_{0}<1$ we have that every single virus generates statistically less than
one new virus, so a permanent infection is not possible and the infected
state does not exist. We also note that the case explored in the present
paper, and so the expression (\ref{19}), is more general than recent
estimations for $R_{0}$ where the possibility of a distributed intracellular
period was also taken into account (Heffernan and Wahl, 2006).

We will now explore the stability of the equilibrium states found. For this
purpose, we will use the usual linear-stability analysis, so we introduce $%
X(t)=X_{\text{eq}}+\delta X(t)$ and $V(t)=V_{\text{eq}}+\delta V(t)$.
Inserting these definitions into (\ref{11}) and linearizing about the
equilibrium state we obtain the following system for the perturbations 
\begin{eqnarray}
\frac{\text{d}\delta X(t)}{\text{d}t} &=&-\beta V_{eq}\delta X(t)-\beta
X_{eq}\delta V(t)-\int_{0}^{t}\delta X(t-t^{\prime })\Psi _{X}(t^{\prime })%
\text{d}t^{\prime }  \nonumber \\
&&+\beta X_{eq}\int_{0}^{t}\delta V(t-t^{\prime })\Psi _{X}(t^{\prime
})t^{\prime }e^{-\beta V_{eq}t^{\prime }}\text{d}t^{\prime }  \nonumber \\
\frac{\text{d}\delta V(t)}{\text{d}t} &=&-\beta X_{eq}\delta V(t)-\beta
V_{eq}\delta X(t)+\beta V_{eq}\int_{0}^{t}\delta X(t-t^{\prime })\phi
(t^{\prime })\Phi _{Y}(t^{\prime })\text{d}t^{\prime }  \nonumber \\
&&+\beta X_{eq}\int_{0}^{t}\delta V(t-t^{\prime })\phi (t^{\prime })\Phi
_{Y}(t^{\prime })\text{d}t^{\prime }  \label{24} \\
&&-\int_{0}^{t}\delta V(t-t^{\prime })\Psi _{V}(t^{\prime })\text{d}%
t^{\prime }+\beta V_{eq}\int_{0}^{t}\delta X(t-t^{\prime })\Psi
_{V}(t^{\prime })t^{\prime }e^{-\beta X_{eq}t^{\prime }}\text{d}t^{\prime }.
\nonumber
\end{eqnarray}
Since this system is now linear, we can propose for the perturbations
exponential solutions of the form $e^{\mu t}$ to get the characteristic
equation

\begin{eqnarray}
0 &=&\left( \mu +\beta X_{eq}+\left[ \Psi _{X}\right] _{\mu }\right) \left(
\mu +\beta X_{eq}-\beta X_{eq}\left[ \phi \Phi _{Y}\right] _{\mu }+\left[
\Psi _{X}\right] _{\mu }\right)  \nonumber \\
&&-\beta ^{2}X_{eq}V_{eq}\left( 1-\frac{\text{d}\left[ \Psi _{X}\right]
_{\mu }}{\text{d}\mu }\right) \left( 1-\left[ \phi \Phi _{Y}\right] _{\mu }-%
\frac{\text{d}\left[ \Psi _{V}\right] _{\mu }}{\text{d}\mu }\right) ,
\label{26}
\end{eqnarray}
where we define $[f]_{\mu }\equiv \int e^{-\mu t}f(t)$d$t$ in accordance
with the notation used above for the Laplace transform.

\textit{a) Infection-free equilibrium state}

First we analyze the stability of the trivial state corresponding to the
absence of viruses. Introducing (\ref{17}) into (\ref{26}) we obtain 
\begin{equation}
1=\beta X_{eq}\left[ \Phi _{V}\right] _{\mu +\beta X_{eq}}\left[ \phi \Phi
_{Y}\right] _{\mu }.  \label{29}
\end{equation}
From (\ref{29}), it is easy to find the necessary condition for the
transition from stability to instability. In the BMVD it is known that the
condition $R_{0}\gtrless 1$ determines the stability of the infected-free
state. From (\ref{29}), it is possible to prove that, in general, this
condition holds for any choice of the PDF's. The right hand side in that
equation is a monotonically decreasing positive function of $\mu $ and takes
the value $R_{0}$ at $\mu =0$. Then, if $R_{0}>1$ both curves always
intersect at a single point for a positive value of $\mu $, which is nothing
but the sufficient condition for the state to be unstable, independently of
the PDF's considered. If $R_{0}<1$ both curves always intersect at a single
point but now for a negative value of $\mu $. In this case the
infection-free equilibrium state is linearly stable and infection dies out.

\textit{b) Infected equilibrium state}

Using (\ref{18}), the characteristic equation (\ref{26}) for the infected
state becomes extremely complicated to treat, and it makes impossible to
determine analytically the stability of the infected state. However, we can
still deduce the behavior of this state by imposing some conditions to
prevent the system from behaving unrealistically. First, we mention again
that the infected state does not exist for $R_{0}<1$, so we only need to
study the case $R_{0}>1$. Second, we can rewrite the first equation in (\ref
{18}), using (\ref{19}) and the definition of the Laplace transform, as 
\begin{equation}
\left[ \varphi _{V}\right] _{\beta X_{eq}}=\frac{\left[ \phi \Phi
_{Y}\right] _{\mu }-1}{\left[ \phi \Phi _{Y}\right] _{\mu }}.  \label{18new}
\end{equation}
Then, we conclude that there is only one possible positive solution for $%
X_{eq}$, as the left hand side of this equation is a monotonically
decreasing function of $X_{eq}$. From that, similar arguments can be applied
to the third equation in (\ref{18}), so it follows that the solution for $%
V_{eq}$ is unique too. As a whole, we have that the infected state is always
unique. This, together with the unstability of the non-infected state for $%
R_{0}>1$, allows us to conclude that the infected state cannot be an
unstable node or a saddle point, as it would imply that for some initial
conditions the system would grow without control towards the state $%
X\rightarrow \infty $ and/or $V\rightarrow \infty $. This unbounded behavior
is not possible in our system. Then, the only possibility is that the
infected state is stable for $R_{0}>1$.

The derivations presented in this Section show that the introduction of
distributed delays does not modify the stability conditions of the BMVD.
Although our mesoscopic model (\ref{11}) is much more general that the
original version (\ref{1}), we find that the condition $R_{0}\gtrless 1$ is
always the one that determine the stability of the two possible equilibrium
states. Note also that the condition to have an infected state of
coexistence between viruses and cells ($R_{0}>1$) can be interpreted as a
threshold value for the contact rate 
\begin{equation}
\beta >\frac{1}{\lambda \tau _{X}\left[ \int_{0}^{\infty }e^{-\beta \lambda
\tau _{_{X}}t}\Phi _{V}(t)\text{d}t\right] \left[ \int_{0}^{\infty }\phi
(t)\Phi _{Y}(t)\text{d}t\right] }.
\end{equation}

\section{The BMVD with a delayed eclipse phase}

We have presented a general model which takes into account distributed
delays for the cellular death and the eclipse phase. However, the
application of the general case requires knowing all the temporal
distributions considered, which is not always possible at practice. Then, it
can be useful to study some specific and simpler cases which have a special
interest for application purposes.

First, we consider the case where no age-distributed effects are introduced
in the death process i.e. the probability of death is independent of the age
of the cells. This corresponds to the situation used in the BMVD, which in
our integrodifferential model is recovered by assuming $\varphi _{X}$, $%
\varphi _{Y}$, $\varphi _{V}$ as exponentially decaying functions ($\varphi
_{X}(t)=\delta e^{-\delta t}$, $\varphi _{Y}(t)=ae^{-at}$, $\varphi
_{V}(t)=ue^{-ut}$). For the eclipse phase, we can assume that when a cell is
infected, it takes a fixed constant time $\tau $ until the first virion is
released and after that, virions are continuously released at a constant
rate $k$. The delay $\tau $ is the time necessary to inject the viral core
into the cell and make its genetic machinery start the reproduction process.
So that, the function $\phi (t)$ in our model will be taken as a step
function $\phi (t)=kH(t-\tau )$,where $H()$ is the Heaviside function.

This specific example has been studied by some authors before (Herz et al,
1996; Tam, 1999; Culshaw and Ruan, 2000), so we can compare the predictions
from our model with those previous approaches. Replacing the distribution
functions $\varphi _{i}(t),$ $\phi (t)$ into the general model (\ref{11}) we
obtain 
\begin{eqnarray}
\frac{\text{d}X}{\text{d}t} &=&\lambda -\delta X-\beta XV  \nonumber \\
\frac{\text{d}Y}{\text{d}t} &=&\beta XV-aY  \nonumber \\
\frac{\text{d}V}{\text{d}t} &=&\int_{\tau }^{t}\beta X(t-t^{\prime
})V(t-t^{\prime })ke^{-at^{\prime }}\text{d}t^{\prime }-\beta XV-uV.
\label{19b}
\end{eqnarray}

In the equation for $V(t)$, the expression $\beta X(t-t^{\prime
})V(t-t^{\prime })$ represents those cells that became infected at time $%
t-t^{\prime }$. So, the new virions appeared are equal to that expression
multiplied by the rate $k$ and by the probability $e^{-at^{\prime }}$ that
the infected cells have survived from time $t-t^{\prime }$ to $t$. The
expression of $R_{0}$ that one obtains for this case, from (\ref{19}), is 
\begin{equation}
R_{0} = \frac{\beta \lambda}{\delta u} \left( \frac{k}{a}e^{-a\tau
}-1\right) .  \label{19c}
\end{equation}

Note that the system (\ref{19b}) is apparently different to the previous
models proposed before for the analysis of a delayed eclipse phase (Herz et
al., 1996; Tam, 1999; Culshaw and Ruan, 2000). In those works a delayed term 
$\beta X(t-\tau )V(t-\tau )$ was introduced \textit{ad hoc} in the evolution
equation for $Y(t)$: 
\begin{eqnarray}
\frac{\text{d}X}{\text{d}t} &=&\lambda -\delta X-\beta XV  \nonumber \\
\frac{\text{d}Y}{\text{d}t} &=&\beta X(t-\tau )V(t-\tau )e^{-a\tau }-aY 
\nonumber \\
\frac{\text{d}V}{\text{d}t} &=&kY-\beta XV-uV.  \label{19d}
\end{eqnarray}
However, it is easy to see that the value of $R_{0}$ for this model is
exactly the expression (\ref{19c}), and the equilibrium states coincide with
those found from our model too. Actually, both models represent the same
underlying process except for one subtle detail. In the model (\ref{19d}),
the fraction of cells $\beta X(t-\tau )V(t-\tau )$ are considered as
infected cells only after the time delay $\tau $. But during the period from 
$t-\tau $ to $\tau $ these cells 'disappear', it is, they do not enter
neither in the equation for $Y$ nor in those for $X$ or $V$. Instead, in our
model the cells become $Y$ cells at time $t-\tau $ and they start releasing
the new virions at time $t$, so our approach is phenomenologically more
correct. Regarding the dynamics of both models, the only difference between (%
\ref{19b}) and (\ref{19d}) will be in the solution for $Y(t)$: the value
predicted by the model (\ref{19d}) will be always below the real one, as
some infected cells are not being counted.

\section{The effect of age-distributed times for cellular death}

Now we try to study a more realistic case according to the experimental data
available in the literature. We will consider that the eclipse phase follows
the same dynamics as that in Section 4. But the death times are now assumed
to follow Gamma distributions, which are quite standard curves used for
fitting experimental data to cellular death times (see for example the
recent work by Hawkins et. al. (2007)). Hence, in this case we will use 
\begin{equation}
\phi (t)=kH(t-\tau )\qquad \varphi _{i}(t)=\frac{t^{\alpha _{i}-1}e^{-t/\tau
_{i}^{*}}}{\left( \tau _{i}^{*}\right) ^{\alpha _{i}}\Gamma (\alpha _{i})}
\label{20}
\end{equation}
for $i=X,Y,V$, where $\Gamma ($\textperiodcentered $)$ denotes the gamma
function and $\alpha _{i}$ and $\tau _{i}^{*}$ are the characteristic
parameters of the Gamma distribution for mortality, with the average
lifetime given by $\tau _{i}=\tau _{i}^{*}\alpha _{i}$.

Inserting these distributions into (\ref{19}) the basic reproductive ratio $%
R_{0}$ reads 
\begin{equation}
R_{0}=\frac{\left( 1+\beta \lambda \tau _{X}\tau _{V}^{*}\right) ^{\alpha
_{V}}-1}{\left( 1+\beta \lambda \tau _{X}\tau _{V}^{*}\right) ^{\alpha _{V}}}%
k\tau _{Y}^{*}e^{-\tau /\tau _{Y}^{*}}\sum_{j=0}^{\alpha _{Y}-1}\left[ \frac{%
\alpha _{Y}-j}{j!}\left( \frac{\tau }{\tau _{Y}^{*}}\right) ^{j}\right]
\label{21}
\end{equation}
for $\alpha _{Y}$ integer. From (\ref{21}), it follows that the influence of
distributed death ages could be important for the value of $R_{0}$ and, as a
result, it strongly modifies the value of the virus load at equilibrium.
This effect is represented in Figure 2, which shows the numerical solution $%
V(t)$ obtained from the model (\ref{11}) for different values of the
parameter $\alpha $ (for simplicity we define $\alpha \equiv \alpha
_{X}=\alpha _{Y}=\alpha _{V}$). For $\alpha =1$ we recover the case where
the death probabilities are exponentially distributed, it is, the prediction
by the BMVD. In the three curves shown, the average lifetimes for the three
species are kept the same. It allows us to compare properly the effects of
the mortality distributions on the virus load dynamics. Two main differences
are observed between the curves in Figure 2. First, note that the virus
loads decrease in time for $t<2$; this is because we have used a delay $\tau
=2$ for the eclipse phase, so only after $t=\tau $ the infected cells start
to release the first virions, and then the virus load increases drastically.
The minimum value observed at $t=2$ is much lower in the case $\alpha =1$.
This is because the BMVD assumes unrealistic high probabilities of death for
the early stage of the infection, an effect which can be corrected by the
Gamma-distributed mortalities used here. This point is of great importance
concerning the probabilities of a primary immune response to successfully
clear the infection. Second, we also find important differences between the
maximum virus loads reached at equilibrium; for the parameters used in
Figure 2, the final virus load for $\alpha =1$ is approximately 10-fold
higher than in the case $\alpha =3$.

Therefore, we conclude that the BMVD underestimates the virus loads in the
early stages of the infection and overestimates the peak of the virus load,
if compared with the case of distributed mortalities considered here. In
consequence, it turns out that we need to know with some detail the life
cycle of viruses and cells to obtain an accurate picture of the infection
dynamics.

\section{Application to phage-bacteria interactions}

The interaction between phages and bacteria can be described as two
consecutive steps: adsorption and reproduction (Mc Grath and Sinder, 2007).
Adsorption consists in a collision between phage and bacteria resulting in a
group, called infected bacteria, constituted by the bacteria and the phage
attached to its membrane. The second step begins when the phage inoculates
its genetic material into the host bacteria and begins to replicate it. From
this time onwards the number of new viruses increases inside the bacteria,
stopping when the bacteria bursts at the end of the latent period.
Basically, the main difference between this situation and those explored in
the previous Sections is that for phages the eclipse phase finishes with a
lytic process that involves the death of the infected cell. In terms of the
model presented here, this idea can be introduced simply by choosing the
appropriate form for the function $\phi (t)$.

Here we deal with the reproduction process, which is known to produce a
characteristic one-step growth curve $V(t)$ for virulent phages. Let us
consider that at $t=0$ the phage inoculates its genome and all the bacteria
become infected instantaneously, with $Y(t=0)=Y(0)$. Then, we can define $%
J_{V}(t)=Y(0)\phi (t)$ as the rate of viruses released at time $t$,
following the same notation as in the Appendix (see Equation (\ref{5}) and
the comments below). As all the cells are assumed to be already infected at $%
t=0$, the infection process for $t>0$ can be obviated. We can thus take $%
\Omega _{V}=1$ in Equation (\ref{5}) to obtain 
\begin{equation}
V(t)=V(0)\Phi _{V}(t)+\int_{0}^{t}Y(0)\phi (t-t^{\prime })\Phi
_{V}(t^{\prime })\text{d}t^{\prime },  \label{vosg}
\end{equation}
which constitutes our theoretical model for the osg curve. If the osg is
known from experiments, the function $\phi (t)$ can be determined by fitting
that curve to some function and applying 
\begin{equation}
\phi (t)=\frac{1}{Y(0)}\left( \frac{\text{d}V}{\text{d}t}%
+\int_{0}^{t}V(t-t^{\prime })\Psi _{V}(t^{\prime })\text{d}t^{\prime
}\right) _{\text{osg}}  \label{phipre}
\end{equation}
which comes directly from the solution of (\ref{vosg}). However, the result (%
\ref{phipre}) can only be applied if we know the function $\Psi _{V}$, which
is related to the mortality distribution $\varphi _{V}$ according to (\ref
{12}). At practice, the probability of death for the viruses is usually
considered very small in the time scale of the experiments, so it can be
neglected. In that case, $\Psi _{V}\approx 0$ and then we find that $\phi
(t) $ becomes proportional to the derivative of the one-step growth (osg)
curve 
\begin{equation}
\phi (t)=\frac{1}{Y(0)}\left( \frac{\text{d}V}{\text{d}t}\right) _{\text{osg}%
}.  \label{fitfi}
\end{equation}
For fitting the one-step growth $V(t)$, some authors have considered before
a piecewise function composed by three segments (You et al., 2002;{\Large \ }%
Hadas et al., 1997). Continuous functions have been proposed too, for
example error functions (Rabinovitch et al., 1999) or logistic-like
functions (Fort and M\'{e}ndez, 2002; Alvarez et al., 2007). For these three
cases one finds that the corresponding expressions for $\phi (t)$ are those
shown in Table 1. We have written there the functions in terms of the
parameters $r$, $\tau $ and $V_{\infty }$. For the sake of completeness, we
also show the relation between these parameters and the eclipse time, the
rise rate and the burst size, which are commonly used in experimental works
to characterize the osg curve (a proper definition of these is provided in
Figure 3).

In Figure 4 we show with symbols the experimental results for one-step
growth of phage T7 on \textit{E. coli} BL21 grown at different rates (You et
al., 2002), while the specific values obtained from the adjustment in each
case are detailed in Table 2. The solid curves in Figure 4 represent the
fitting of the experimental results to the logistic-like function,
exhibiting a good agreement. The segments (dotted lines) and the error
function (dashed lines) fittings are also showed in the plot; in the latter,
the coincidence with the logistic-like case is so high that both curves are
almost indistinguishable.

From each one of the fittings the corresponding expression for $\phi (t)$
has been estimated. The comparison between them is shown in Figure 5, where
we plot for simplicity only one of the three cases presented in Figure 4
(the two cases non-shown exhibit a very similar behavior). We observe that
for the 'error' and the 'logistic-like' cases, peaked $\phi (t)$ functions
with very similar characteristics are obtained. The 'segments' case, in
turn, leads to a discontinuous expression for $\phi (t)$ which slightly
differs from the other two. So, we can conclude that the 'segments' fitting
gives a poorer estimate for the behavior of $\phi (t)$ and this can
influence the final value for $R_0$.

We note that in this specific application for phages a new definition of $%
R_{0}$ is necessary, as can be seen by inspecting (\ref{19}). To this end,
we must find the equilibrium states of the system 
\begin{eqnarray}
\frac{\text{d}X}{\text{d}t} &=&-\beta X(t)V(t)  \nonumber \\
\frac{\text{d}V}{\text{d}t} &=&-\beta X(t)V(t)+\int_{0}^{t}\beta
X(t-t^{\prime })V(t-t^{\prime })k(t^{\prime })\phi (t^{\prime })\text{d}%
t^{\prime }
\end{eqnarray}
and their stability. Introducing $X(t)=X_{\text{eq}}+\delta X(t)$ and $%
V(t)=V_{\text{eq}}+\delta V(t)$ and linearizing about the equilibrium states 
$(X_{\text{eq}},0)$ and $(0,V_{\text{eq}})$ one can check that the basic
reproductive ratio 
\begin{equation}
R_{0}\equiv \int_{0}^{\infty }k(t)\phi (t)\text{d}t
\end{equation}
must be higher than 1 for a successful phage growth. Making use of (\ref
{fitfi}) 
\begin{equation}
R_{0}=\frac{1}{Y(0)}\int_{0}^{\infty }\left( \frac{\text{d}V}{\text{d}t}%
\right) _{\text{osg}}\text{d}t=\frac{\left[ V_{\infty }-V(0)\right] _{\text{%
osg}}}{Y(0)}
\end{equation}
which is the burst size. This result simply demonstrates that in the case of
phage-bacteria interactions the burst size plays the role of a basic
reproductive ratio (the infection is successful only for $R_{0}>1$).

\section{Conclusions}

In the present paper, we have derived a generalization of the basic model of
virus dynamics by considering a more accurate life cycle for viruses and
cells which includes distributed delays for mortality and the eclipse phase.
As a result, we have showed how the infection dynamics gets modified. As
discussed above, our main motivation has been to present a rigorous approach
to this problem, as many times delays have been introduced in this kind of
models just by intuitive or \textit{ad hoc} arguments. For this reason, we
have provided here a mesoscopic derivation based on explicit balance
equations that provide a very accurate description of the underlying
dynamical process. In our approach, the life cycles are implemented in a
probabilistic way by the distributions functions $\varphi _{X}$, $\varphi
_{Y}$, $\varphi _{V}$ and $\phi $. Then, this is a very powerful and general
formalism, provided that one has the data necessary to evaluate these
functions.

We have carried out a formal analysis of the equilibrium states and their
stability. Furthermore, we have illustrated how the model works for some
simple situations of interest. Specifically, for phage-bacteria interactions
we have been able to provide analytical expressions that may serve to
estimate the function $\phi (t)$ from the one-step growth curves.

In short, the main conclusions obtained from our study are the following:

\textit{i)} The mesoscopic formalism presented here allows to reduce the
BMVD of 3-species to only 2 species ($X$ and $V$). Then, albeit our model
requires a more complex mathematical treatment, this simplification can be
an interesting advantage.

\textit{ii)} We have formally proved that the stability diagram of the BMVD
is insensitive to any delays considered. It means that the model has always
an equilibrium infected-free state which becomes unstable for $R_{0}>1$,
which is exactly the same condition necessary for the existence of a stable
infected state. This generalizes similar results previously found (Culshaw
and Ruan, 2000; Nelson and Perelson, 2002; Wang et. al., 2007) that reached
the same conclusions for more specific cases.

\textit{iii)} The reproductive ratio $R_{0}$ and the virus loads are in
general very sensitive to the distributed mortalities considered. It proves
one needs to know in detail the cellular life cycles, in special for
viruses, to describe the infection process. Models based on exponentially
distributed death rates will provide non-accurate results for the virus
dynamics, which can lead to wrong predictions concerning the success or
failure of an immune response or vaccination.

\textit{iv}) For phage-bacteria interactions, we have found that fittings of
the one-step growth based on logistic-like and error functions yield very
similar expressions for $\phi (t)$. From the analysis shown here, it is not
possible to determine which one of them is more accurate. Anyway, it is
clear that both cases give better and more realistic estimates for $\phi (t)$
and $R_0$ than fittings based on three segments.

In short, we have found that introducing age-distributed processes in the
BMVD may strongly modify the dynamics of viral infections. These corrections
can be of great interest when the effects of an immune response are also
considered in the model. Then, the dynamics of the model is expected to
become richer (as happens without delays, too) and the role of the cellular
life cycles could be more dramatic. Specifically, we expect that
age-distributed processes can be able to induce new dynamical patterns as
periodicity or chaos, in the line of recent works on this field (Liu, 1997;
Buric et. al., 2001; Canabarro et. al., 2004; Wang et. al., 2007). We will
address these ideas in a forthcoming paper.

\section*{Appendix. Derivation of the model.}

We have introduced $\varphi _{X}(t)$, $\varphi _{Y}(t)$ and $\varphi _{V}(t)$
as the mortality PDF's (see Figure 1). So that, the probability that a
target cell which 'was born' at time $t=0$ has not died yet at time $t$ is
given by $\Phi _{X}(t)$ (hereafter we will refer to it as the ''survival
probability'') according to 
\begin{equation}
\Phi _{X}(t)=1-\int_{0}^{t}\varphi _{X}(t^{\prime })\text{d}t^{\prime },
\label{2}
\end{equation}
and analogous arguments hold for $\Phi _{Y}(t)$ and $\Phi _{V}(t)$.

Then, we can write the balance equation for the population densities as 
\begin{eqnarray}
X(t) &=&X(0)\Phi _{X}(t)\Omega _{X}(0,t)+\int_{0}^{t}J_{X}(t-t^{\prime
})\Phi _{X}(t^{\prime })\Omega _{X}(t-t^{\prime },t)\text{d}t^{\prime }
\label{3} \\
Y(t) &=&Y(0)\Phi _{Y}(t)+\int_{0}^{t}J_{Y}(t-t^{\prime })\Phi _{Y}(t^{\prime
})\text{d}t^{\prime }  \label{4} \\
V(t) &=&V(0)\Phi _{V}(t)\Omega _{V}(0,t)+\int_{0}^{t}J_{V}(t-t^{\prime
})\Phi _{V}(t^{\prime })\Omega _{V}(t-t^{\prime },t)\text{d}t^{\prime },
\label{5}
\end{eqnarray}
where $J_{X}(t)$ represents the density of particles of species $X$ appeared
at time $t$, with equivalent definitions for $J_{Y}(t)$ and $J_{V}(t)$. The
function $\Omega _{X}(t-t^{\prime },t^{\prime })$ is the probability that a
particle $X$ do not become infected during the time interval $(t-t^{\prime
},t)$, while $\Omega _{V}(t-t^{\prime },t)$ is the probability that a virus
has not been adsorbed by a cell during that interval. So that, the balance
equation (\ref{3}) simply says that the density of particles $X$ at time $t$
is given by the initial density of particles $X(0)$ not infected yet and
still alive, plus those target cells appeared at any time so far, provided
they have neither died nor become infected yet. The meaning of Equations (%
\ref{4}-\ref{5}) follow analogous arguments.

Regarding the functions $\Omega $, their explicit form can be found in the
following way. For $\Omega _{X}$ we take 
\begin{equation}
\Omega _{X}(t-t^{\prime },t)=\exp \left[ -\int_{t-t^{\prime }}^{t^{\prime
}}\beta V(t^{\prime \prime })\text{d}t^{\prime \prime }\right]  \label{5b}
\end{equation}
which corresponds to the solution of the infection equation d$X/$d$t=-\beta
XV$. As the infection is considered independent on the other processes
(death and production of new cells by the host), the solution of that ODE
within the interval $(t-t^{\prime },t)$ gives us a proper definition for the
probability $\Omega _{X}(t-t^{\prime },t)$. Similarly, from d$V/$d$t=-\beta
XV$ we can write 
\begin{equation}
\Omega _{V}(t-t^{\prime },t)=\exp \left[ -\int_{t-t^{\prime }}^{t^{\prime
}}\beta X(t^{\prime \prime })\text{d}t^{\prime \prime }\right] .  \label{5c}
\end{equation}

The validity of the expressions (\ref{5b}-\ref{5c}) can be demonstrated from
more rigorous arguments using the age-structured models by Vlad and Ross
(2002). In fact, our model (\ref{3}-\ref{5}) can be seen as a particular
case of the very general model by Yadav and Horsthemke (2006), which was in
turn based on the original work (Vlad and Ross, 2002). Accordingly, we will
follow the formalism in (Yadav and Horsthemke, 2006) to derive our model.

First, we differentiate the system (\ref{3}-\ref{5}) with respect to $t$: 
\begin{eqnarray}
\frac{\text{d}X}{\text{d}t} &=&-X(0)\Omega _{X}(0,t)\left[ \varphi
_{X}(t)+\beta V(t)\Phi _{X}(t)\right] +J_{X}(t)  \nonumber \\
&&-\int_{0}^{t}J_{X}(t-t^{\prime })\varphi _{X}(t^{\prime })\Omega
_{X}(t-t^{\prime },t)\text{d}t^{\prime }  \nonumber \\
&&-\beta V(t)\int_{0}^{t}J_{X}(t-t^{\prime })\Phi _{X}(t^{\prime })\Omega
_{X}(t-t^{\prime },t)\text{d}t^{\prime }  \label{5d} \\
\frac{\text{d}Y}{\text{d}t} &=&-Y(0)\varphi
_{Y}(t)+J_{Y}(t)-\int_{0}^{t}J_{Y}(t-t^{\prime })\varphi _{Y}(t^{\prime })%
\text{d}t^{\prime }  \label{5e} \\
\frac{\text{d}V}{\text{d}t} &=&-V(0)\Omega _{V}(0,t)\left[ \varphi
_{V}(t)+\beta X(t)\Phi _{V}(t)\right] +J_{V}(t)  \nonumber \\
&&-\int_{0}^{t}J_{V}(t-t^{\prime })\varphi _{V}(t^{\prime })\Omega
_{V}(t-t^{\prime },t)\text{d}t^{\prime }  \nonumber \\
&&-\beta X(t)\int_{0}^{t}J_{V}(t-t^{\prime })\Phi _{V}(t^{\prime })\Omega
_{V}(t-t^{\prime },t)\text{d}t^{\prime }  \label{5f}
\end{eqnarray}
Then, we introduce (\ref{3}) and (\ref{5}) into (\ref{5d}) and (\ref{5f}),
respectively, so we obtain 
\begin{eqnarray}
\frac{\text{d}X}{\text{d}t} &=&-X(0)\Omega _{X}(0,t)\varphi
_{X}(t)+J_{X}(t)-\beta X(t)V(t)  \nonumber \\
&&-\int_{0}^{t}J_{X}(t-t^{\prime })\varphi _{X}(t^{\prime })\Omega
_{X}(t-t^{\prime },t)\text{d}t^{\prime }  \label{5g} \\
\frac{\text{d}Y}{\text{d}t} &=&-Y(0)\varphi
_{Y}(t)+J_{Y}(t)-\int_{0}^{t}J_{Y}(t-t^{\prime })\varphi _{Y}(t^{\prime })%
\text{d}t^{\prime }  \label{5h} \\
\frac{\text{d}V}{\text{d}t} &=&-V(0)\Omega _{V}(0,t)\varphi
_{V}(t)+J_{V}(t)-\beta X(t)V(t)  \nonumber \\
&&-\int_{0}^{t}J_{V}(t-t^{\prime })\varphi _{V}(t^{\prime })\Omega
_{V}(t-t^{\prime },t)\text{d}t^{\prime }.  \label{5i}
\end{eqnarray}

On the other side, we divide (\ref{3}) by $\Omega _{X}(0,t)$ and transform
that equation to the Laplace domain (again, we denote the Laplace transform
of a function by the brackets $[$\textperiodcentered $]_{s}$ with the
conjugate variable $s$). After some easy algebra, it can be written as 
\begin{equation}
\frac{s[\varphi _{X}]_{s}}{1-[\varphi _{X}]_{s}}\left[ \frac{X}{\Omega
_{X}(0,t)}\right] _{s}=X(0)[\varphi _{X}]_{s}+[\varphi _{X}]_{s}\left[
J_{X}\Omega _{X}\right] _{s}.  \label{5j}
\end{equation}
Finally, introducing the inverse Laplace transform of (\ref{5j}) into (\ref
{5g}), the evolution equation for the species $X$ reads 
\begin{equation}
\frac{\text{d}X}{\text{d}t}=J_{X}(t)-\beta XV-\int_{0}^{t}X(t-t^{\prime
})\Psi _{X}(t^{\prime })\Omega _{X}(t-t^{\prime },t)\text{d}t^{\prime },
\label{5l}
\end{equation}
where $\Psi _{X}$ is defined in the Laplace domain by (\ref{12}). For the
species $Y$ and $V$ we can use exactly the same derivation, so that the
Equations (\ref{5h},\ref{5i}) turn into 
\begin{eqnarray}
\frac{\text{d}Y}{\text{d}t} &=&J_{Y}(t)-\int_{0}^{t}Y(t-t^{\prime })\Psi
_{Y}(t^{\prime })\text{d}t^{\prime }  \label{5n} \\
\frac{\text{d}V}{\text{d}t} &=&J_{V}(t)-\beta XV-\int_{0}^{t}V(t-t^{\prime
})\Psi _{V}(t^{\prime })\Omega _{V}(t-t^{\prime },t)\text{d}t^{\prime }
\label{5o}
\end{eqnarray}
with $\Psi _{Y}$, $\Psi _{V}$ defined implicitly in (\ref{12}).

Hence, we have obtained the general evolution equations (\ref{5l}-\ref{5o})
for the model. However, note that we still need to give expressions for the
densities $J_{i}$. From Equation (\ref{1}), the number of new target cells
appearing at any given time can be expressed as 
\begin{equation}
J_{X}(t)=\lambda ,  \label{6}
\end{equation}
Similarly, the density of infected cells appearing at time $t$ is given by
the expression 
\begin{equation}
J_{Y}(t)=\beta X(t)V(t).  \label{7}
\end{equation}

Finally, the new viruses appeared at time $t$ are given by the function $%
\phi (t)$ (see Figure 1) applied over those cell which were infected at any
previous time $t-t^{\prime }$, provided they have not died yet. This allows
us to write 
\begin{equation}
J_{V}(t)=\int_{0}^{t}J_{Y}(t-t^{\prime })\phi (t^{\prime })\Phi
_{Y}(t^{\prime })\text{d}t^{\prime }.  \label{8}
\end{equation}

Once we have the explicit expressions for $J_{X}$, $J_{Y}$ and $J_{V}$, our
model takes the final form (\ref{11}): 
\begin{eqnarray*}
\frac{\text{d}X}{\text{d}t} &=&\lambda -\beta XV-\int_{0}^{t}X(t-t^{\prime
})\Psi _{X}(t^{\prime })\Omega _{X}(t-t^{\prime },t)\text{d}t^{\prime } \\
\frac{\text{d}Y}{\text{d}t} &=&\beta XV-\int_{0}^{t}Y(t-t^{\prime })\Psi
_{Y}(t^{\prime })\text{d}t^{\prime } \\
\frac{\text{d}V}{\text{d}t} &=&\int_{0}^{t}\beta X(t-t^{\prime
})V(t-t^{\prime })k(t^{\prime })\phi (t^{\prime })\Phi _{Y}(t^{\prime })%
\text{d}t^{\prime } \\
&&-\beta XV-\int_{0}^{t}V(t-t^{\prime })\Psi _{V}(t^{\prime })\Omega
_{V}(t-t^{\prime },t)\text{d}t^{\prime }.
\end{eqnarray*}

\section*{Acknowledgments.}

This research has been partially supported by the Generalitat de Catalunya
by the grant 2006-BP-A-10060 (DC), and by Grants Nos. FIS 2006-12296-C02-01,
SGR 2005-00087 (VM) and EPSRC EP/D03115X/1 (SF and VM).

\section*{References}

Allegrini, P., Aquino, G., Grigolini, P., Palatella, L., Rosa, A., 2003.
Generalized master equation via aging continuous-time random walks. Phys.
Rev. E 68, 056123.

Allegrini, P., Bologna, M., Grigolini, P., West, B.J., 2007.
Fluctuation-Dissipation Theorem for Event-Dominated Processes. Phys. Rev.
Lett. 99, 010603.

Alvarez, L.J., Thomen, P., Makushok, T., Chatenay, D., 2007. Propagation of
fluorescent viruses in growing plaques. Biotechnol. Bioeng. 96, 615-621.

Anderson, R.M., May, R.M., 1991. Infectious diseases of humans. Oxford
University Press.

Banks, H.T., Bortz, D.M., Holte, S.E., 2003. Incorporation of variability
into the modeling of viral delays in HIV infection dynamics. Math. Biosci.
183, 63-91.

Berkowitz, B., Scher, H., Silliman, S.E., 2000. Anomalous transport in
laboratory-scale, heterogeneous porous media.Water Resour. Res. 36, 149158.

Bonhoeffer, S., May, R.M., Shaw, G.M., Nowak, M.A., 1997. Virus dynamics and
drug therapy. Proc. Natl. Acad. Sci. 94, 6971-6976.

Buric, N., Mudrinic, M., Vasovic, N., 2001. Time delay in a basic model of
the immune response. Chaos Soliton. Fract. 12, 483-489.

Canabarro, A.A., Gl\'{e}ria, I.M., Lyra, M.L., 2004. Periodic solutions and
chaos in a non-linear model for the delayed cellular immune response.
Physica A 342, 234-241.

Culshaw, R.V., Ruan, S. A., 2000. Delay-differential equation model of HIV
infection of CD4 T-cells. Math. Biosci. 165, 27-39.

Emmanuel, S., Berkowitz, B., 2007. Continuous-time random walks and heat
transfer in porous media. Transport in Porous Media 67, 413-430.

Fedotov, S., Iomin, A., 2007. Migration and proliferation dichotomy in
tumor-cell invasion. Phys. Rev. Lett. 98, 118101.

Fort, J., M\'{e}ndez, V., 2002. Time-delayed spread of viruses in growing
plaques. Phys. Rev. Lett. 89, 178101.

Hadas, H., Einav, M., Fishov, I., Zaritsky, A., 1997. Bacteriophage T4
development depends on the physiology of its host Escherichia coli.
Microbiology 143, 179-185.

Hawkins, E.D., Turner, M.L., Dowling, M.R., van Gend, C., Hodgkin, P.D.,
2007. A model of immune regulation as a consequence of randomized lymphocyte
division and death times. Proc. Natl. Acad. Sci. 104, 5032-5037.

Heffernan, J.M., Wahl, L.M., 2006. Improving estimates of the basic
reproductive ratio: Using both the mean and the dispersal of transition
times. Theor. Pop. Biol. 70, 135-145.

Helmstetter, A., Sornette, D., 2002. Diffusion of epicenters of earthquake
aftershocks, Omoris law, and generalized continuous-time random walk models.
Phys. Rev. E 66, 061104.

Herz, A.V.M., Bonhoeffer, S., Anderson, R.M., May, R.M., Nowak, M.A., 1996.
Viral dynamics in vivo: Limitations on estimates of intracellular delay and
virus decay. Proc. Natl. Acad. Sci. 93, 7247-7251.

Li, D., Wanbiao, M., 1999. Asymptotic properties of a HIV-1 infection model
with time delay. J. Math. Anal. Appl. 335, 683-691.

Liu, W., 1997. Nonlinear Oscillations in Models of Immune Responses to
Persistent Viruses. Theor. Pop. Biol. 52, 224-230.

Lloyd, A.L., 2001. The dependence of viral parameter estimates on the
assumed viral life cycle: limitations of studies of viral load data. Proc.
R. Soc. Lond. B 268, 847-854.

Masoliver, J., Montero, M., Perell\'{o}, J. Weiss, G.H., 2006. The CTRW in
finance: Direct and inverse problems with some generalizations and
extensions. Physica A 379, 151-167.

Mc Grath, S., van Sinder, D., 2007. Bacteriophage: Genetics and Molecular
Biology. Caister Academic Press, UK.

Mittler, J.E., Sulzer, B., Neumann, A.U., Perelson, A.S., 1998. Influence of
delayed viral production on viral dynamics in HIV-1 infected patients. Math.
Biosci. 152, 143-163.

Nelson, P.W., Perelson, A.S., 2002. Mathematical analysis of delay
differential equation models of HIV-1 infection. Math. Biosci. 179, 73-94.

Nowak, M.A., Bangham, C.R.M., 1996. Population dynamics of immune responses
to persistent viruses. Science 272, 74-79.

Nowak, M.A., May, R.M., 2000. Virus dynamics. Mathematical principles of
immunology and virology. Oxford University Press.

Perelson, A.S., 2002. Modelling viral and immune system dynamics. Nature
Rev. Immun. 2, 28-36.

Rabinovitch A., Hadas, H., Einav, M., Melamed, Z., Zaritsky, A., 1999. Model
for bacteriophage T4 development in Escherichia coli. J. Bacteriol. 181,
1677-1683.

Rebenshtok, A., Barkai, E., 2007. Distribution of time-averaged observables
for weak ergodicity breaking. Phys. Rev. Lett. 99, 210601.

Tam, J., 1999. Delay effect in a model for virus replication. IMA J. Math.
Appl. Med. Biol. 16, 29-37.

Vlad, M.O., Ross, J., 2002. Systematic derivation of reaction-diffusion
equations with distributed delays and relations to fractional
reaction-diffusion equations and hyperbolic transport equations: Application
to the theory of Neolithic transition. Phys. Rev. E 66, 061908.

Wang, K., Wang, W., Pang, H., Liu, X., 2007. Complex dynamic behavior in a
viral model with delayed immune response. Physica D 226, 197-208.

Wearing, H.J., Rohani, P., Keeling, M.J., 2005. Appropriate models for the
management of infectious diseases. PLoS Med. 2, 0621-0627

Wodarz, D., Klenerman, P., Nowak, M.A., 1998. Dynamics of cytotoxic
T-lymphocyte exhaustion. Proc. Roy. Soc. London B 265, 191-203.

Wodarz, D., Nowak, M.A., 1999. Specific therapy regimes could lead to
long-term immunological control of HIV. Proc. Natl. Acad. Sci. USA 96,
14464-14469.

Wodarz, D., 2006. Ecological and evolutionary principles in immunology.
Ecol. Lett. 9, 694-705.

Yadav, A., Horsthemke, W., 2006. Kinetic equations for reaction-subdiffusion
systems: Derivation and stability analysis. Phys. Rev. E 74, 066118.

You, L., Suthers, P.F., Yin, J., 2002. Effects of Escherichia coli
physiology on growth of phage T7 in vivo and in silico. J. Bacteriol. 184,
1888-1894. \newpage

Figure Captions

Figure 1. Scheme of the BMVD model with distributed delays for mortality and
the eclipse phase.

Figure 2. Virus loads obtained numerically from the general model (\ref{11})
for the case of Gamma-distributed death times. In the legend we show the
values of the parameter $\alpha =\alpha _{X}=\alpha _{Y}=\alpha _{V}$ used.
The other parameters used are $\beta =0.02,$ $k=50$, $\tau =2$, $\tau
_{X}^{*}=10/\alpha $, $\tau _{Y}^{*}=10/\alpha $, $\tau _{V}^{*}=10/\alpha $.

Figure 3. Definition of the eclipse time, the rise rate and the burst size
in terms of the one-step growth curve.

Figure 4. One-step growth for phage T7 inside \textit{E. coli}. The lines
shown represent the fittings from the functions in Table 1 (solid lines
represent the logistic-like fitting, dashed lines correspond to the error
function, and the dotted lines the segments fitting). Symbols correspond to
experimental results, obtained from You et al. (2002). Circles, triangles
and diamonds represent the osg-curve for the host growing at 0.7, 1.0 and
1.2 doublings/h, respectively.

Figure 5. Comparison between the function $\phi (t)$ predicted from the
three different fittings proposed in Table 1. Results shown correspond to
the case of growing at 0.7 doublings per hour shown in Figure 3. The solid,
dashed and dotted lines correspond to the predictions from the
logistic-like, error function and segments, respectively.

\newpage

Table 1

Characteristics of the three functions proposed for fitting the osg curve,
with their explicit expressions for the eclipse time, the rise rate and the
burst size. From (\ref{fitfi}), the estimations for $\phi (t)$ are also
shown.

\bigskip 
\begin{tabular}[b]{llllll}
\hline
& $V(t)$ & $\phi (t)$ & $
\begin{array}{l}
\text{Eclipse} \\ 
\text{time}
\end{array}
\text{ }$ & $
\begin{array}{l}
\text{Rise} \\ 
\text{rate}
\end{array}
\text{ }$ & $
\begin{array}{l}
\text{Burst} \\ 
\text{size}
\end{array}
\text{ }$ \\ \hline\hline
$
\begin{array}{l}
\text{Three} \\ 
\text{segments:}
\end{array}
\text{ }$ & $\left\{ 
\begin{array}{l}
0;\text{ }t<\tau \\ 
r(t-\tau );\text{ }\tau <t<\tau +\frac{V_{\infty }}{r} \\ 
V_{\infty };\text{ }t>\tau +\frac{V_{\infty }}{r}
\end{array}
\right. $ & $
\begin{array}{l}
H\left( t-\tau \right) \\ 
-H\left( t-\tau -\frac{V_{\infty }}{r}\right)
\end{array}
$ & $\tau $ & $r$ & $V_{\infty }$ \\ \hline
$\text{Error function:}$ & $\frac{V_{\infty }}{2}\left[ 1-\text{erf}\left( 
\frac{\tau -t}{4}r\sqrt{\pi }\right) \right] $ & $\frac{r}{4}e^{-r^{2}\pi
(t-\tau )^{2}/16}$ & $\tau -2/r$ & $rV_{\infty }/4$ & $V_{\infty }$ \\ \hline
$\text{Logistic-like:}$ & $\frac{V_{\infty }}{1+e^{-r(t-\tau )}}$ & $\frac{%
re^{-r(t-\tau )}}{\left[ 1+e^{-r(t-\tau )}\right] ^{2}}$ & $\tau -2/r$ & $%
rV_{\infty }/4$ & $V_{\infty }$ \\ \hline
\end{tabular}

Table 2

Values of the parameters obtained from the fittings shown in Figure 3

\[
\begin{tabular}{llll}
\hline
& Three segments & Error function & Logistic-like \\ \hline\hline
Burst size\textbf{\ }(PFU/ml) & 35.9$\pm 1.5$ ($\circ $) & 36$\pm 1$ ($\circ 
$) & 37$\pm 1$ ($\circ $) \\ 
\textbf{\ }$\ \ \ \ \ \ \ \ \ \ \ \ \ \ \ $ & 35.7$\pm 0.9$ ($\triangle $) & 
37.8$\pm 0.6$ ($\triangle $) & 38.3$\pm 0.6$ ($\triangle $) \\ 
\textbf{\ }$\ \ \ \ \ \ \ \ \ \ \ \ \ \ \ $ & 75.2$\pm 2.5$ ($\Diamond $) & 
75$\pm 25$ ($\Diamond $) & 76$\pm 25$ ($\Diamond $) \\ \hline
Rise rate\textbf{\ (}PFU ml$^{-1}\min^{-1})$ & 1.5$\pm 0.1$ ($\circ $) & 1.7$%
\pm 0.2$ ($\circ $) & 1.8$\pm 0.2$ ($\circ $) \\ 
& 3.4$\pm 0.4$ ($\triangle $) & 3.6$\pm 0.2$ ($\triangle $) & 3.8$\pm 0.2$ ($%
\triangle $) \\ 
& 4.9$\pm 0.2$ ($\Diamond $) & 5.9$\pm 0.65$ ($\Diamond $) & 6.3$\pm 0.75$ ($%
\Diamond $) \\ \hline
eclipse time\textbf{\ }(min) & 24.1$\pm 0.9$ ($\circ $) & 25.6$\pm 0.4$ ($%
\circ $) & 26.3$\pm 1.6$ ($\circ $) \\ 
& 21.1$\pm 0.5$ ($\triangle $) & 21.4$\pm 0.1$ ($\triangle $) & 21.7$\pm 0.4$
($\triangle $) \\ 
& 17.9$\pm 0.3$ ($\Diamond $) & 19.1$\pm 0.25$ ($\Diamond $) & 19.5$\pm 0.85$
($\Diamond $) \\ \hline
\end{tabular}
\]

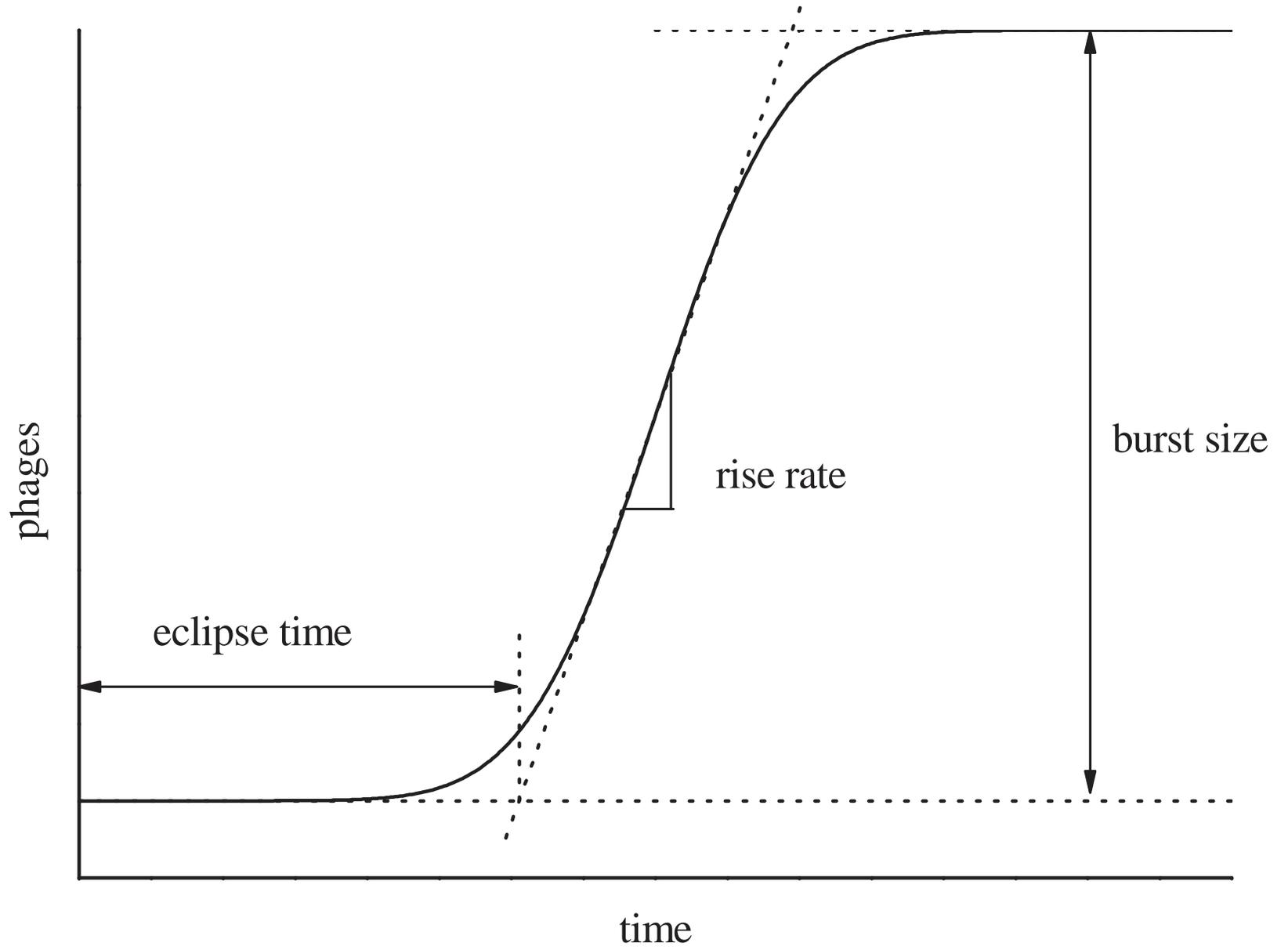
\end{document}